\documentclass[12pt]{article}
\usepackage{amsfonts}
\usepackage{color}
\usepackage[active]{srcltx}

\begin{document}

\def\ba{\begin{eqnarray}}
\def\ea{\end{eqnarray}}
\def\w{\wedge}

\begin{titlepage}
\title{ \bf  Spinor coupling to the weak Poincare gauge theory of gravity in three dimensions}
\author{ \"{O}zcan Sert\footnote{\tt sertoz@itu.edu.tr} , Muzaffer Adak\footnote{\tt madak@pau.edu.tr}\\ \\
 {\small Department of Physics, Faculty of Arts and Sciences, Pamukkale University}\\
 {\small 20017 Denizli, Turkey }   }
 \vskip 1cm
\date{ 21.November.2012 {\it file TorsionDiracFinal.tex}}

\maketitle

\begin{abstract}

\noindent The Dirac lagrangian is minimally coupled to the most
general $R+T+T^2$-type lagrangian in (1+2)-dimensions. The field
equations are obtained from the total lagrangian by a variational
principle. The space-time torsion is calculated algebraically in
terms of the Dirac condensate plus coupling coefficients. A family
of circularly symmetric rotating exact solutions which is
asymptotically $AdS_3$ is obtained. Finally BTZ-like solutions are
discussed.

\vskip 1cm



\end{abstract}
\end{titlepage}

\section{Introduction}

 \noindent
Although it is well known that general relativity is a classically
trivial theory in three dimensions, the proposition of
topologically massive gravity of Deser, Jackiw and Tempelton
\cite{deser} made it non-trivial and thus increased considerably
theoretical interest in 3D gravity. In the meantime the discovery
of Banados-Teitelboim-Zanelli (BTZ) black holes \cite{teitelboim}
enhanced 3D gravity efforts, see e.g.
\cite{Horne1989}-\cite{Nazaroglu2011} and references therein. The
motivations for those investigations can be listed briefly as the
study of the properties of the quantum fields in curved spacetimes
\cite{Birrell1982}, inflation \cite{Guth1981} and the dS/CFT
correspondence \cite{Strominger2001},\cite{Bousso2002}.

\medskip
\noindent On the other hand, the non-Riemannian formulation is
another approach to be followed to obtain a dynamical 3D theory of
gravity. There is a plenty of literature on 3D gravity with
torsion. The first possibility along this route is the
Einstein-Cartan theory. Nevertheless it is nondynamic in the
absence of matter. Thus it is amended by the inclusion of
Chern-Simons term. Then Mielke and Baekler generalized the
topological massive gauge model of gravity by adding a new
translational Chern-Simons term to the standard (rotational) one
\cite{mielke1991}. This generalization with or without matter
attracted a lot of attention in the literature, see for example
\cite{Garcia2003}-\cite{Sert2012} and references therein.

\medskip
\noindent On the contrary, the number of the published works on
the spinor coupled 3D gravity model with/without torsion is much
less, to our knowledge, \cite{Sert2012}-\cite{Dereli2010}. Our
initial aim is to fill in this gap. Nevertheless, first time in
the literature we investigate 3D gravity which is formulated in
terms of {\it the most} general non-propagating torsion. That is,
we write a lagrangian in the form of $R+T+T^2$ which is also
called the weak Poincare gauge theory of gravity. Thus our gravity
lagrangian contains six parameters, $a,\lambda,k_1,k_2,k_3,b$.
When the Dirac spinor is minimally coupled to it, $k_2$ disappears
and one of $k_1$ or $k_3$ can be dropped without loss of
generality. Also $b$ gives contributions to both the bare
cosmological constant and the mass of Dirac spinor.

\medskip
\noindent The paper is organized as follows. Since we will be
using the coordinate independent exterior forms, in Section
\ref{sec:Notation} we introduce our notations and conventions. In
Section \ref{sec:Fieldeqns}, after we couple minimally the Dirac
lagrangian to the gravitational lagrangian, we obtain the FIRST
and SECOND field equations and the Dirac equation by varying the
total lagrangian with respect to the coframe, the connection and
the adjoint of Dirac spinor, respectively. Before closing this
section we solve torsion from the SECOND equation and insert the
findings to the FIRST equation. After that, in Subsection
\ref{sec:riemanntheory} we reduce our theory to a Riemannian one.
Section \ref{sec:cirsymsol} starts with a circularly symmetric and
rotating metric ansatz. Then we write explicitly the Dirac
equation and cast the FIRST equation as five coupled differential
equations. In order to see whether there is an exact solution to
our model, in Subsection \ref{susec:alpgam} we restrict ourselves
to a special case, $\alpha = \gamma$ by tracing the technique in
\cite{Dereli2010}. Here we obtain a family of solutions which goes
to $AdS_3$ as $r \rightarrow \infty$. In Subsection
\ref{susec:h1ovf} we consider BTZ-like solutions and do find one,
but only for the case of vanishing Dirac condensate.

\section{Mathematical preliminaries} \label{sec:Notation}

\noindent We specify the space-time geometry by a triplet $ \left (
M, g, \nabla \right )$ where $M$ is a 3-dimensional differentiable
manifold equipped with a metric tensor
 \ba
   g = \eta_{ab} e^a \otimes e^b
 \ea
of signature  $(-,+,+)$. $e^a$ is an orthonormal co-frame dual to
the frame vectors $X_a$, that is $e^a(X_b) \equiv \iota_b e^a=
\delta^{a}_{b}$ where $\iota_b := \iota_{X_b}$ denotes the
interior product. A metric compatible connection $\nabla$ can be
specified in terms of connection 1-forms ${\omega^a}_b$ satisfying
$\omega_{ba} = -\omega_{ab}$. Then the Cartan structure equations
 \ba
    de^a + {\omega^a}_b \w e^b = T^a \, , \\
     d{\omega^a}_b + {\omega^a}_c \w {\omega^c}_b = {R^a}_b
 \ea
define the space-time torsion 2-forms $T^a$ and curvature 2-forms
${R^a}_b$, respectively. Here $d$ denotes the exterior derivative
and $\w$ the wedge product. We fix the orientation of space-time
by choosing the volume 3-form ${}^*1 = e^0 \wedge e^1 \wedge e^3$
where ${}^*$ is the Hodge star map. In three dimensional
space-times with Lorentz signature for any $p$-form ${}^{**}=-1$.
We will use the abbreviation $e^{ab \cdots} := e^a \w e^b \w
\cdots $. It is possible to decompose the connection 1-forms in a
unique way as
 \ba
   {\omega^a}_b = \widetilde{\omega}^a_{\;\;b} +{K^a}_b
   \label{eq:FullConnec}
 \ea
where $\widetilde{\omega}^a_{\;\;b}$ are the zero-torsion
Levi-Civita connection 1-forms satisfying
 \ba
     de^a + \widetilde{\omega}^a_{\;\;b} \w e^b = 0
 \ea
and ${K^a}_b$ are the contortion 1-forms satisfying
 \ba
    {K^a}_b \w e^b = T^a \, . \label{eq:Contor1}
 \ea
Correspondingly, the full curvature 2-form is decomposed as
Riemannian part plus torsional contributions:
 \ba
    {R^a}_b = \widetilde{R}^a_{\;\;b} + \widetilde{D}{K^a}_b + {K^a}_c \w {K^c}_b
 \ea
where $\widetilde{R}^a_{\;\;b}$ is the Riemannian curvature 2-form
and
$$
\widetilde{D}{K^a}_b = d{K^a}_b + \widetilde{\omega}^a_{\;\;c} \w
{K^c}_b - \widetilde{\omega}^c_{\;\;b} \w {K^a}_c \,.
$$
As seen above, we label the Riemannian quantities by a tilde.

\medskip
\noindent We are using the formalism of Clifford algebra
$\mathcal{C}\ell_{1,2}$-valued exterior forms.
$\mathcal{C}\ell_{1,2}$ algebra is generated by the relation among
the orthonormal basis $\{\gamma_0,\gamma_1,\gamma_2\}$
 \begin{equation}
    \gamma^a \gamma^b + \gamma^b \gamma^a = 2\eta^{ab} \, .
 \end{equation}
One particular representation of the $\gamma^\alpha$'s is given by the following Dirac matrices
 \begin{eqnarray}
   \gamma_0 = \left(\begin{array}{cc}
                0 & 1 \\
                -1 & 0
              \end{array}\right) \, , \;
   \gamma_1 = \left(\begin{array}{cc}
                0 & 1 \\
                1 & 0
              \end{array}\right) \, , \;
   \gamma_2 = \left(\begin{array}{cc}
                1 & 0 \\
                0 & -1
              \end{array}\right) \, .
 \end{eqnarray}
In this case a Dirac spinor $\Psi$ can be represented by a
2-component column matrix. Thus we write explicitly the covariant
exterior derivative of $\Psi$, its Dirac conjugate and the
curvature of the spinor bundle, respectively,
 \ba
  D\Psi = d\Psi + \frac{1}{2}  \sigma_{ab} \Psi \omega^{ab} \; , \quad
  D\overline{\Psi} = d\overline{\Psi} - \frac{1}{2} \overline{\Psi} \sigma_{ab}  \omega^{ab} \; , \quad
  D^2\Psi = \frac{1}{2}R^{ab} \sigma_{ab} \Psi
 \ea
where $\sigma_{ab}:= \frac{1}{4}[\gamma_a , \gamma_b]=\frac{1}{2}
\epsilon_{abc} \gamma^c$ are the generators of the Lorentz group.
Overline figures the Dirac adjoint, $\overline{\Psi}:= \Psi^\dag
\gamma_0$. We frequently make use of the identity
 \ba
     \gamma_c \sigma_{ab} + \sigma_{ab} \gamma_c = \epsilon_{abc} \, .
 \ea

\section{The Weak Poincare gauge theory of gravity} \label{sec:Fieldeqns}

\noindent The field equations of our model are obtained by varying the action
 \ba
   I[e^a,\omega^{ab},\overline{\Psi}] = \int_M \left ( L_G + L_D  \right )
 \ea
where $L_G$ signifies the gravitational lagrangian density 3-form
 \ba
  L_G &=& \frac{a}{2} R_{ab} \w {}^*e^{ab} + \lambda {}^*1 +\frac{k_1}{2} T^a \w
  {}^*T_a \nonumber \\
  & &+ \frac{k_2}{2} \mathcal{V} \w {}^*\mathcal{V} + \frac{k_3}{2} \mathcal{A} \w {}^*\mathcal{A}
  + \frac{b}{2} T^a \w e_a
 \ea
and $L_D$ denotes the (hermitian) Dirac lagrangian density 3-form
 \ba
   L_D = \frac{i}{2} \left ( \overline{\Psi}\, {}^*\gamma \w D \Psi - D\overline{\Psi} \w \, {}^*\gamma \Psi \right )
   + i m \overline{\Psi} \Psi \, {}^*1 \label{eq:DirLag}
 \ea
with the definitions $\mathcal{V} = \iota_a T^a$ and $\mathcal{A}=
T^a \w e_a$. Here the gravitational constants $a,k_1,k_2,k_3$,
mass $m$ and the Dirac field $\Psi$ have the dimension of
$length^{-1}$,  the gravitational constant $b$ has the dimension
of $length^{-2}$, and the cosmological constant $\lambda$ has the
dimension of $length^{-3}$. When all $k_1,k_2,k_3,b$ coefficients
are zero, it corresponds the well-known Einstein-Cartan-Dirac
theory with cosmological constant. The hermiticity of the
lagrangian (\ref{eq:DirLag}) leads to a charge current which
admits the usual probabilistic interpretation. $L_G$ is the most
general gravity lagrangian with non-propagating torsion in three
dimensions. It is also called the weak Poincare gauge theory of
gravity in three dimensions. We remind that a term containing an
odd number of the Hodge star has even parity and its coefficient
is {\it scalar} and that with even Hodge star has odd parity and
its factor is {\it pseudoscalar}. Correspondingly, we notice that
$a,k_1,k_2,k_3,m$ are scalar, but $b$ is pseudoscalar.
$\frac{b}{2} T^a \w e_a$ is known as the {\it translational
Chern-Simons term} which corresponds the usual (rotational)
Chern-Simons 3-form, $ (1/2) ({\omega^a}_b \w d{\omega^b}_a +
(2/3) {\omega^a}_b \w {\omega^b}_c \w {\omega^c}_a)$, for the
curvature \cite{mielke1991}.

\medskip
\noindent We obtain the field equations via independent variations
with respect to $e^a, \omega^{ab}, \overline{\Psi}$. Thus
$e^a$-variation yields the FIRST equation
 \ba
    -\frac{a}{2} \epsilon_{abc} R^{bc} - \lambda {}^*e_a - b T_a  & & \nonumber \\
    - \frac{k_1}{2} \left[ 2D^*T_a + \iota_a(T^b \w ^*T_b) - 2(\iota_aT^b)\w ^*T_b \right] & & \nonumber\\
     + \frac{k_2}{2} \left[ 2D(\iota_a{}^*\mathcal{V}) - \iota_a(\mathcal{V} \w ^*\mathcal{V})
    - 2(\iota_aT^b)\w (\iota_b{}^*\mathcal{V})\right] & & \nonumber\\
     - \frac{k_3}{2} \left[ 2D(e_a \w {}^*\mathcal{A}) + \iota_a(\mathcal{A} \w ^*\mathcal{A})
     - 2(\iota_aT^b)\w (e_b \w {}^*\mathcal{A})\right] &=& \tau_a \, , \label{eq:FldEqCoframe}
 \ea
$\omega^{ab}$-variation yields the SECOND equation
  \ba
    -\frac{a}{2} \epsilon_{abc} T^c + \frac{b}{2} e_{ab}
    + \frac{k_1}{2} (e_a \w {}^*T_b - e_b \w {}^*T_a) & & \nonumber \\
    - \frac{k_2}{2} (e_a \w \iota_b {}^* \mathcal{V} - e_b \w \iota_a {}^*\mathcal{V})
    + k_3 e_{ab} \w {}^* \mathcal{A}  &=& \Sigma_{ab}  \, , \label{eq:FldEqConnec}
 \ea
and $\overline{\Psi}$-variation yields the Dirac equation
 \ba
     {}^*\gamma \w (D-\frac{1}{2}\mathcal{V})\Psi + m \Psi {}^*1
     =0 \, , \label{eq:DirEqn}
 \ea
where $ \Sigma_{ab}= - \mathcal{S} e_{ab}$ is the Dirac angular momentum 2-form with
$\mathcal{S} := \frac{i}{4} \overline{\Psi} \Psi$ and
$\tau_a$ is the Dirac energy-momentum 2-form
 \ba
  \tau_a = \frac{i}{2}  {}^*e_{ba} \w \left[ \overline{\Psi} \gamma^b
  (D\Psi) - (D\overline{\Psi}) \gamma^b \Psi \right] + im\overline{\Psi} \Psi
  {}^*e_a \, .
 \ea
For future convenience by using the Dirac equation
(\ref{eq:DirEqn}) and its conjugate
$(D-\frac{1}{2}\mathcal{V})\overline{\Psi} \w {}^*\gamma -m
\overline{\Psi} {}^*1=0$ we rewrite the Dirac energy-momentum
2-form as
 \ba
    \tau_a &=& -\frac{i}{2} \left[ \overline{\Psi} \gamma_b (D_a\Psi) - (D_a\overline{\Psi}) \gamma_b \Psi
    \right] {}^*e^b \nonumber \\
   &=& -\frac{i}{2} \left[ \overline{\Psi} \gamma_b (\partial_a \Psi) - (\partial_a\overline{\Psi}) \gamma_b \Psi
    \right] {}^*e^b + \mathcal{S} \omega_{bc,a} e^{bc}
    \label{eq:DirEnMom}
 \ea
where $D_a := \iota_a D$, $\partial_a := \iota_a d$ and
$\omega_{bc,a} := \iota_a \omega_{bc}$.

\medskip
\noindent Now we solve the SECOND equation (\ref{eq:FldEqConnec})
for torsion
 \ba
    T^a = \mathcal{P} \, {}^*e^a \quad \mbox{where} \; \; \mathcal{P}=\frac{2\mathcal{S}+b}{-a+2(k_1+3k_3)} \,
    . \label{eq:SolOfTorsion}
 \ea
Then we calculate $\mathcal{V}=0$ and $\mathcal{A}=3\mathcal{P}
{}^*1$. By substituting these results into the FIRST equation
(\ref{eq:FldEqCoframe}) we obtain
 \ba
    \frac{a}{2}\epsilon_{abc}R^{bc} + (k_1+3k_3) e_a \w d
    \mathcal{P} + [\lambda + b\mathcal{P} -
    \frac{1}{2}(k_1+3k_3)\mathcal{P}^2] {}^*e_a + \tau_a =0 \, .
    \label{eq:first}
 \ea
The manner in which $(k_1+3k_3)$ appears in the equations
(\ref{eq:SolOfTorsion}) and (\ref{eq:first}) makes it clear that
one can set $k_1=0$ or $k_3=0$ without loss of generality.
Instead, we redefine $k_1+3k_3 =c$. For later use, we also note
that substitution of (\ref{eq:SolOfTorsion}) into
(\ref{eq:Contor1}) yields the following expression for the
contortion
 \ba
    K_{ab} = -\frac{\mathcal{P}}{2} {}^*e_{ab} \, .
    \label{eq:contor2}
 \ea

\subsection{Reduction to a Riemannian theory}
\label{sec:riemanntheory}

\noindent To gain physical insight on the coupling parameters and
torsion, we reformulate the theory in terms of Riemannian
quantities. Firstly we decompose the concerned quantities by using
(\ref{eq:SolOfTorsion}) and (\ref{eq:contor2}) repeatedly,
 \ba
    R_{ab} \w {}^*e^{ab} &=& \widetilde{R}_{ab} \w {}^*e^{ab} +
    \frac{3}{2} \mathcal{P}^2 \, {}^*1 + mod(d) \, , \\
  T^a \w {}^*T_a &=& -3\mathcal{P}^2 \, {}^*1 \, ,\\
  \mathcal{A} \w {}^*\mathcal{A} &=& -9 \mathcal{P}^2 \, {}^*1 \, ,\\
  T^a \w e_a &=& 3\mathcal{P} {}^*1 \, ,\\
  D\Psi &=& \widetilde{D}\Psi + \frac{\mathcal{P}}{4} \gamma \Psi \, ,\\
  D\overline{\Psi} &=& \widetilde{D} \overline{\Psi} -
  \frac{\mathcal{P}}{4} \overline{\Psi} \gamma \, .
 \ea
Here since $mod(d):=\left( \widetilde{D}K_{ab} \right) \w
{}^*e^{ab} = d\left( K_{ab} \w {}^*e^{ab} \right)$ is an exact
form it can be discarded. Also $\widetilde{D}\Psi$ is defined as $
\widetilde{D}\Psi= d\Psi + \frac{1}{2} \sigma_{ab} \Psi
\widetilde{\omega}^{ab}$, and similarly
$\widetilde{D}\overline{\Psi}$ is. When we insert all those
findings into the total lagrangian, $L=L_G + L_D$, we obtain a new
Riemannian lagrangian which is equivalent to the weak Poincare
gauge theory of gravity,
 \ba
    \widetilde{L} &=& \frac{a}{2} \widetilde{R}_{ab} \w {}^*e^{ab} +
    \rho_a \w T^a + \left[ \lambda + \frac{3b^2}{4(2c -a)} \right]
    {}^*1 \nonumber \\
    & & + \frac{i}{2} \left ( \overline{\Psi}\, {}^*\gamma \w \widetilde{D} \Psi
    - \widetilde{D}\overline{\Psi} \w \, {}^*\gamma \Psi \right )
   + i \left[m + \frac{3(\mathcal{S}+b)}{4(2c-a)}\right] \overline{\Psi} \Psi \,
   {}^*1 \, , \label{eq:riemanlag}
 \ea
where $\rho_a$ is a lagrange multiplier 1-form constraining
torsion to zero. As seen above, pseudoscalar coupling coefficient
$b$ shifts the bare cosmological constant and the mass of the
Dirac particle. In fact, the Dirac field gains mass through
torsional interactions.

\section{Circularly symmetric rotating solutions} \label{sec:cirsymsol}

\noindent We consider the metric
 \ba
    g = - f^2(r) dt^2 + h^2(r) dr^2 + r^2 \left( w(r) dt + d \phi \right)^2
 \ea
in plane polar coordinates $(t, r, \phi)$. Here the metric
function $w(r)$ is concerned with rotation. We use the notation
and the techniques introduced in \cite{Dereli2010}. The following
choice of the orthonormal basis 1-forms
 \ba
   e^0 = f(r) dt \, , \quad e^1 = h(r) dr \, , \quad e^2 = r (w(r) dt + d \phi) ,
 \ea
leads to the Levi-Civita connection 1-forms
 \ba
    \widetilde{\omega}^{0}_{\; \;1} = \alpha e^0 - \frac{\beta}{2} e^2 \, , \quad
    \widetilde{\omega}^{0}_{\; \; 2} = -\frac{\beta}{2} e^1 \, , \quad
    \widetilde{\omega}^{1}_{\; \; 2} = -\frac{\beta}{2} e^0 - \gamma
    e^2 \label{eq:levicivita}
 \ea
where we defined
 \ba
    \alpha = \frac{f^{\prime}}{f h } \, , \quad
    \beta = \frac{r w^{\prime}}{f h } \, , \quad
    \gamma = \frac{1}{r h} \, . \label{eq:AlpBetGam}
 \ea
Here prime denotes the derivative with respect to $r$. Then we
write explicitly the full connection 1-forms with the substitution
of (\ref{eq:contor2}) and (\ref{eq:levicivita}) into the equation
(\ref{eq:FullConnec})
 \ba
    \omega_{01} = -\alpha e^0 + \frac{\beta - \mathcal{P}}{2} e^2 \, , \;\;
    \omega_{02} = \frac{\beta + \mathcal{P}}{2} e^1 \, , \;\;
    \omega_{12} = - \frac{\beta + \mathcal{P}}{2} e^0 - \gamma e^2 \, .
 \ea
Under the assumption of $\Psi=\Psi(r)$ we calculate the curvature
2-forms
 \ba
   & &{R^0}_1 = \left( -\frac{\alpha'}{h}-\alpha^2 + \frac{3\beta^2}{4} + \frac{\mathcal{P}^2}{4} \right)e^{01}
   + \left( \frac{\mathcal{P}'-\beta'}{2h} - \beta \gamma \right)e^{12} \, , \nonumber \\
   & &{R^0}_2 =  \left(- \alpha \gamma + \frac{\mathcal{P}^2-\beta^2}{4}  \right)e^{02} \, , \nonumber \\
   & &{R^1}_2 = \left( \frac{\mathcal{P}'+\beta'}{2h} + \beta \gamma \right)e^{01}
   + \left( -\frac{\gamma'}{h}-\gamma^2 + \frac{\mathcal{P}^2-\beta^2}{4} \right)e^{12} \, .
 \ea
The next operation is to write down the Dirac equation
(\ref{eq:DirEqn}) and its adjoint
 \ba
  & & \frac{\Psi'}{h} = -\left[ \frac{\alpha + \gamma}{2}
   + \left(\frac{\beta}{4} + \frac{3\mathcal{P}}{4} + m \right) \gamma_1\right] \Psi \, ,\label{eq:DirEqn1} \\
  & & \frac{\overline{\Psi}'}{h} = -\overline{\Psi}\left[ \frac{\alpha + \gamma}{2}
   - \left(\frac{\beta}{4} + \frac{3\mathcal{P}}{4} + m \right)
   \gamma_1\right] \, . \label{eq:DirEqn2}
 \ea
Now we can calculate explicitly the Dirac energy-momentum 2-forms by using the equation (\ref{eq:DirEnMom})
 \ba
   & & \tau_0 = -2\alpha \mathcal{S} e^{01} - \mathcal{S}(\beta + \mathcal{P}) e^{12} \, , \nonumber \\
   & & \tau_1 = -(2 \mathcal{P} \mathcal{S} + 4m\mathcal{S}) e^{02} \, , \nonumber \\
   & & \tau_2 = \mathcal{S} (\beta - \mathcal{P}) e^{01} -2\gamma \mathcal{S} e^{12}\, .
 \ea
Then the FIRST equation (\ref{eq:first}) turns out to be the
following set of the coupled ordinary differential equations
 \ba
    \frac{\beta'}{2h} + \frac{(a-2c)\mathcal{P}'}{2ah} + \beta
    \gamma - \frac{2\alpha \mathcal{S}}{a} =0 \, , \label{eq:first1}\\
   \frac{\beta'}{2h} - \frac{(a-2c)\mathcal{P}'}{2ah} + \beta
    \gamma + \frac{2\gamma \mathcal{S}}{a} =0 \, ,\label{eq:first2}\\
    -\frac{\alpha'}{h}-\alpha^2 + \frac{3\beta^2}{4} +
    \frac{(a-2c)\mathcal{P}^2}{4a} + \frac{\mathcal{S}(\beta -
    \mathcal{P})+b\mathcal{P}+\lambda}{a} =0 \, ,\label{eq:first3}\\
  -\frac{\gamma'}{h}-\gamma^2 - \frac{\beta^2}{4} +
    \frac{(a-2c)\mathcal{P}^2}{4a} - \frac{\mathcal{S}(\beta +
    \mathcal{P})-b\mathcal{P}-\lambda}{a} =0 \, ,\label{eq:first4}\\
  -\alpha \gamma - \frac{\beta^2}{4} +
    \frac{(a-2c)\mathcal{P}^2}{4a} + \frac{2\mathcal{P}\mathcal{S}+b\mathcal{P}+ 4m\mathcal{S}+\lambda}{a}
    =0 \, . \label{eq:first5}
 \ea

\subsection{$\alpha(r) = \gamma(r)$ case} \label{susec:alpgam}

\noindent We firstly restrict our attention to those cases for which
 \ba
    \gamma = \alpha = \frac{1}{rh} \, .
 \ea
From the definitions (\ref{eq:AlpBetGam}) it immediately follows
that
 \ba
    f(r) = f_0 r
 \ea
where $f_0$ is a constant. Then, using
(\ref{eq:first1})$\pm$(\ref{eq:first2}) we arrive at
 \ba
    \beta(r) = \frac{\beta_0}{r^2} \, , \quad \quad \mathcal{S}(r) =
    \frac{\mathcal{S}_0}{r^2} \label{eq:DirCondan}
 \ea
where $\beta_0$ and $\mathcal{S}_0$ are integration constants.
But, (\ref{eq:first3})$-$(\ref{eq:first4}) causes a constraint
among them
 \ba
     \beta_0 = -\frac{2}{a} \mathcal{S}_0 \, .
 \ea
Inserting the above results and $\alpha = 1/rh$ into
(\ref{eq:first3})$+$(\ref{eq:first4}) yields a solution for $h(r)$
 \ba
     h(r) = 1/\sqrt{h_0 - AS_0^2/2r^2 + \Lambda r^2}
     \label{eq:sol01}
 \ea
where $h_0$ is an integration constant, $A$ is the shifted
coupling constant and $\Lambda$ is the effective cosmological
constant
 \ba
    A = \frac{4(2a-c)}{a^2(a-2c)} \, , \quad \quad
    \Lambda = \frac{4\lambda (a-2c)-3b^2}{4a(a-2c)} \, .
 \ea
The equation (\ref{eq:first4}) gives a constraint between the
integration constants
 \ba
       h_0 =\frac{4M}{a} \mathcal{S}_0
 \ea
where $M$ is a shifted Dirac mass $M=\frac{4m
(a-2c)-3b}{4(a-2c)}$. Now we calculate $w(r)$ from
(\ref{eq:AlpBetGam}ii)
 \ba
     w(r) = -\sqrt{\frac{8f_0^2}{Aa^2}}
     \arctan\left[\frac{\sqrt{2\Lambda}r^2 + \sqrt{-A \mathcal{S}_0^2
     + 8 M \mathcal{S}_0 r^2/a +2\Lambda r^4}}{\mathcal{S}_0 \sqrt{A}}\right] + w_0
 \ea
where $w_0$ is a constant. Here we notice the consistency
condition $A,\Lambda > 0$. Moreover, if we choose
 \ba
     w_0 = \sqrt{\frac{2\pi^2 f_0^2}{Aa^2}}
 \ea
then as $\mathcal{S}_0 \rightarrow 0$, $w(r)$ goes to zero. These
results have been crosschecked by the computer algebra system,
Reduce \cite{Hearn1993} and its package Excalc \cite{Schrufer}.

\medskip

\noindent Our final job is to work out the Dirac equation. Let us consider a
Dirac spinor field and its Dirac conjugate
 \ba
    \Psi = \left(\begin{array}{c}
             \psi_1(r) \\
             \psi_2(r) \\
           \end{array} \right) \, , \quad \quad \overline{\Psi}:=\Psi^\dag \gamma_0 = \left( \begin{array}{cc}
                                                                   -\psi_2^\star(r) & \psi_1^\star(r) \\
                                                                 \end{array}
                                                                 \right) \label{eq:DirSpinor}
 \ea
where ${}^\star$ denotes complex conjugation and $\psi_1, \psi_2$
are complex functions. Then the equation (\ref{eq:DirEqn1}) reads
in components as follows
 \ba
     \psi_1'=-h\alpha \psi_1 - \frac{\beta +3\mathcal{P}+4m}{4}h\psi_2 \, , \\
  \psi_2'=-h\alpha \psi_2 - \frac{\beta +3\mathcal{P}+4m}{4}h\psi_1 \, .
 \ea
We take the combinations $\psi_\pm = \psi_1 \pm \psi_2$ and write
a decoupled system of equations
 \ba
    \psi_\pm'=-\left( \alpha  \pm \frac{\beta +3\mathcal{P}+4m}{4} \right) h\psi_\pm \, .
 \ea
The explicit solutions to these equations are given by
 \ba
     \psi_\pm (r) = \frac{C_\pm}{r} e^{\mp\left[ -A \varphi(r) + M \theta(r)\right]}
 \ea
where $C_\pm$ are the complex integration constants and
 \ba
     \varphi(r) &=& \frac{a}{4}\int h(r) \mathcal{S}(r) dr \nonumber \\
    &=& \sqrt{\frac{a^2}{8A}}
     \arctan\left[\frac{\sqrt{2\Lambda}r^2 + \sqrt{-A \mathcal{S}_0^2
     + 8M\mathcal{S}_0 r^2/a +2\Lambda r^4}}{\mathcal{S}_0 \sqrt{A}}\right] \, , \\
     \theta(r) &=& \int h(r) dr \nonumber \\
    &=& \frac{1}{\sqrt{4\Lambda}} \ln\left[\frac{ 4M\mathcal{S}_0/a + 2\Lambda r^2
    + \sqrt{2\Lambda(-A\mathcal{S}_0^2 +8M \mathcal{S}_0r^2/a+ 2\Lambda r^4)}}{\sqrt{2A\Lambda\mathcal{S}_0^2
    + 16M^2\mathcal{S}_0^2/a^2}}\right] \, .
 \ea
Thus we can write the components of the Dirac spinor as $\psi_1 =
(\psi_+ + \psi_-)/2$ and $\psi_2 = (\psi_+ - \psi_-)/2$.
Consequently we write down explicitly the Dirac condensate
$\mathcal{S} := \frac{i}{4} \overline{\Psi} \Psi =
\frac{i}{8r^2}(C_-^\star C_+ - C_+^\star C_-)$. By comparing this
with (\ref{eq:DirCondan}ii) we observe
 \ba
   \mathcal{S}_0 = \frac{i}{8}(C_-^\star C_+ - C_+^\star C_-) \, .
 \ea
Here we want to remark that if $C_I$'s, $I=-,+$, are the ordinary
complex numbers (i.e. $C_I C_J = + C_J C_I$, $(C_I C_J)^\star =
C_I^\star C_J^\star$, $C_I^{\star \star}=C_I$) then
$\mathcal{S}_0$ is a real number. Similarly, if $C_I$'s are the
Grassmann complex numbers (i.e. $C_I C_J = - C_J C_I$, $(C_I
C_J)^\star = C_J^\star C_I^\star$, $C_I^{\star \star}=C_I$) then
$\mathcal{S}_0$ is again a real number.

\subsection{$h(r)=1/f(r)$ case} \label{susec:h1ovf}

\noindent We try to find a family of solutions in the form of
$h(r) = 1/f(r)$. By substituting this into the Dirac equation
(\ref{eq:DirEqn1}) with notation (\ref{eq:DirSpinor}) and
$\psi_\pm = \psi_1 \pm \psi_2$ we obtain
 \ba
     \psi_\pm (r) = \frac{C_\pm}{r} e^{\mp \theta(r)}
 \ea
where
 \ba
    \theta (r) = - \int \left(\frac{M}{f(r)} + \frac{j}{2r^2f(r)}  \right) dr \, .
 \ea
Here $j$ is a constant. Then we calculate the Dirac condensate as
 \ba
     \mathcal{S}(r) = \frac{\mathcal{S}_0}{rf(r)}
 \ea
where $\mathcal{S}_0 = \frac{i}{8} (C_-^\star C_+ - C_+^\star
C_-)$. Now if we choose $\mathcal{S}_0=0$, then the set of
equations (\ref{eq:first1})-(\ref{eq:first5}) accepts a family of
solutions as follows
 \ba
    f(r) = \sqrt{\Lambda r^2 - \mathcal{M} + j^2 / r^2} \, , \quad h(r) = 1/ f(r) \, , \quad w(r) =
    j/r^2 \label{eq:BTZ}
 \ea
where $\mathcal{M}$ is an integration constant. This looks like
exactly the same as the very-well known BTZ metric of the General
Relativity.

\section{Conclusion}

 \noindent
We have formulated the Dirac coupled gravity theory with {\it the
most} general non-propagating torsion (the weak Poincare gauge
theory of garvity) in (1+2)-dimensions by using the algebra of
exterior differential forms. We obtained the field equations by a
variational principle. The space-time torsion was calculated
algebraically from the SECOND field equation in terms of the
coupling constants and the quadratic spinor invariant, the
so-called the Dirac condensate. Further, we reformulated the
non-Riemannian theory in terms of Riemannian quantities. Thus we
could gain new interpretations on the coupling coefficients and
the mass of Dirac field.

\medskip

\noindent We then looked for rotating circularly symmetric
solutions, and found a particular class of solutions which is
asymptotically $AdS_3$. These solutions exhibit one singularity at
the origin and two more at the outer region. In order to obtain
the physical meaning of the above singularities, we calculated the
following pair of invariants. The first is the curvature scalar
 \[
    \mathcal{R}= \frac{[3b^2(2a-3c)-6\lambda(a-2c)^2]r^4 +[12b(a-c)-8m(a-2c)^2]\mathcal{S}_0r^2
    + 12 c \mathcal{S}_0^2}{a(a-2c)^2r^4}
 \]
and the second is the quadratic torsion
 \[
   {}^*(T^a \w {}^*T_a) = \frac{3(br^2 + 2
   \mathcal{S}_0)^2}{(a-2c)^2r^4} \, .
 \]
As seen above, although the singularities at outer region are
coordinate singularities,  the singularity at the origin is
essential. Correspondingly, that solution seems to define a black
hole with two horizons. We also remark that if one sets
$\mathcal{S}_0=0$, then both invariants turn out to be constant.

\medskip

\noindent Finally we obtained a BTZ-type solution in the case of
vanishing condensate. Although we searched if the equations
(\ref{eq:first1})-(\ref{eq:first5}) accepted the BTZ solution when
$\mathcal{S}_0 \neq 0$, we were not able to arrive to a definite
answer. This fact, however, does not diminish the novelty of our
solution, because our space-time is still non-Riemannian because
of the non-zero torsion, see the equation (\ref{eq:SolOfTorsion}).
Accordingly, the autoparallel curves of our geometry do not
coincide with the geodesics of metric (\ref{eq:BTZ}). We also
noticed that the coupling parameter $b$ still shifts the mass term
of the Dirac field, see the last parenthesis of
(\ref{eq:riemanlag}). That is, even if the Dirac field was
massless, it would gain mass through the $b$-contained
interactions.

\section*{Acknowledgement}

\noindent We would like thank the anonymous referee for the
enlightening criticisms.

\end{document}